\documentstyle[]{article}
\begin{document}
\overfullrule 0 mm
\language 0
\centerline { \bf{ ONE-DIMENSIONAL MOTION }}
\centerline { \bf{ IN POTENTIAL HOLE }}
\centerline { \bf{OF SOMMERFELD SPHERE }}
\centerline { \bf {IN CLASSICAL
ELECTRODYNAMICS:}}
\centerline { \bf{ INSIDE THE HOLE }}
\vskip 0.5 cm

\centerline {\bf{ Alexander A.  Vlasov}}
\vskip 0.3 cm
\centerline {{  High Energy and Quantum Theory}}
\centerline {{Department of Physics}}
\centerline {{ Moscow State University}}
\centerline {{  Moscow, 119899}}
\centerline {{ Russia}}
\vskip 0.3cm
{\it Equation of motion of Sommerfeld sphere in the
one-dimensional potential hole, produced by two equal charges on some
distance from each other,  is numerically investigated.  Two
types of solutions are found: (i) damping oscillations, (ii)
oscillations without damping (radiationless motion).
Solutions with growing amplitude ("climbing-up-the-wall
solution") for chosen initial conditions were not founded.}

03.50.De
\vskip 0.3 cm
Here we continue our numerical investigation of
one-dimensional motion of Sommerfeld sphere with total charge $Q$,
mechanical mass $m$ and radius $a$ [1].

We consider the one dimensional motion in the symmetrical potential
hole, produced by Coulomb fields of two equal point charges $q$ at
distance $2D$ apart them (Coulomb field has one important
property - it generates the force, acting on the uniformly
charged sphere, of the same value as if the charge of sphere was
concentrated at its center).

For dimensionless variables $y= R/2a,\ \ x=ct/2a,\ \ d=D/2a$
 the  equation of motion of the sphere is

$${d^2 y \over dx^2} =\left(1-({dy\over dx})^2\right)^{3/2}  \cdot k$$
$$ \left[ - \int\limits_{x^{-}}^{x^{+}} dz
{z-1 \over L^2} + \ln {{L^{+}\over L^{-}}} + ({1\over
\beta^2}-1)\ln { {1+\beta \over 1-\beta}} -{2\over \beta}  +
{M\over (y+d)^2} -{M\over (d-y)^2}
 \right] \eqno(1)$$ here    $M=q/Q$ ,
 $$x^{\pm}=1 \pm L^{\pm},\ \
L^{\pm}=y(x)-y(x-x^{\pm}),\ \ L=y(x)-y(x-z),$$
$$ \beta=dy/dx, \
\ \ \ k= {Q^2 \over 2 m c^2 a}.$$

Later on we take $k=1$.

It is useful to compare solutions of (1) with  point
charge motion in the same field, governed by the following
relativistic equation without radiation force:
$${d^2 y \over dx^2} =\left(1-({dy\over dx})^2\right)^{3/2}  \cdot  k
\left[    +{M\over (y+d)^2} -{M\over (d-y)^2}
 \right] \eqno(2)$$

\vskip 0.5 cm {\bf A.}

For chosen initial conditions we have found two types of
solutions of eq. (1):

(i) damping oscillations,

(ii) oscillations without damping (radiationless motion).

Existence of radiationless solutions for Sommerfeld model was
discovered long time ago by Schott [2]. In our case this type of
solutions one can easily obtain from weak-velocity approximation for
eq.(1), when $dy/dx\ll 1,\ \ y\ll d$ and  eq.(1) takes the form
$${d^2 y \over dx^2} = -w^2\cdot y+{4k\over 3}\left[{d  \over
dx}y(x-1) -{d \over dx}y(x) \right] \eqno (3)$$
here $w^2= 4kM/d^3,$

with solution
$$y(x)=A\cos(wx)$$
for $w= 2\pi n,\ \ n=\pm 1,2,...   .$

Solutions with growing amplitude ("climbing-up-the-wall
solution") for chosen initial conditions:
$${dy\over dx}=0 \ \ for \ \ x \leq 0$$
were not founded for wide range  of values of $y(x=0),\ \ M,\ \
d$.

Later on we'll try to search for exotic solutions of Sommerfeld
sphere motion taking another type of initial conditions.

I am glad to thank my colleagues:

P.A.Polyakov - for theoretical discussions;

V.A.Iljina and P.K.Silaev - for assistance rendered during numerical
calculations.

 \centerline {\bf{REFERENCES}}

  \begin{enumerate}
\item Alexander A.Vlasov, physics/9901051.
\item G.A. Schott, Phil.Mag. 15, 752(1933).

\end{enumerate}

 \end{document}